\documentclass[12pt]{iopart}
\usepackage{iopams}  
\usepackage{graphicx}

\def \Poincare  {Poincar\'{e}}

\begin{document}
\title[Turbulence-induced melting of a nonequilibrium vortex crystal]
{Turbulence-induced melting of a nonequilibrium vortex crystal in a
  forced thin fluid film} \author{Prasad Perlekar~$^{1,2}$ and Rahul
  Pandit$^{1,3}$}

\address{$^1$ Centre for Condensed Matter Theory, Department of
  Physics, Indian Institute of
  Science, Bangalore 560012, India\\
  $^2$ Current Address: Department of Mathematics and 
  Computer Science, Technische Universiteit Eindhoven, Eindhoven,
  The Netherlands\\
  $^3$ Also at: Jawaharlal Nehru Centre For Advanced Scientific
  Research, Jakkur, Bangalore, India}
\ead{\mailto{p.perlekar@tue.nl},
  \mailto{rahul@physics.iisc.ernet.in}}

\begin{abstract}
  To develop an understanding of recent experiments on the
  turbulence-induced melting of a periodic array of vortices in a thin
  fluid film, we perform a direct numerical simulation of the
  two-dimensional Navier-Stokes equations forced such that, at low
  Reynolds numbers, the steady state of the film is a square lattice
  of vortices. We find that, as we increase the Reynolds number, this
  lattice undergoes a series of nonequilibrium phase transitions,
  first to a crystal with a different reciprocal lattice and then to a
  sequence of crystals that oscillate in time. Initially the temporal
  oscillations are periodic; this periodic behaviour becomes more and
  more complicated, with increasing Reynolds number, until the film
  enters a spatially disordered nonequilibrium statistical steady that
  is turbulent. We study this sequence of transitions by using
  fluid-dynamics measures, such as the Okubo-Weiss parameter that
  distinguishes between vortical and extensional regions in the flow,
  ideas from nonlinear dynamics, e.g., \Poincare~maps, and theoretical
  methods that have been developed to study the melting of an
  equilibrium crystal or the freezing of a liquid and which lead to a
  natural set of order parameters for the crystalline phases and
  spatial autocorrelation functions that characterise short- and long-range
  order in the turbulent and crystalline phases, respectively.
\end{abstract}


\maketitle

\section{Introduction}

A crystal melts into a disordered liquid when the temperature is
raised beyond the melting point $T_M$; and when the liquid is cooled
below $T_M$ it freezes again. This melting or freezing transition, one
of the most common equilibrium phase transitions, has been studied
extensively; and the Ramakrishnan-Yussouff density-functional
theory~\cite{ram79,hay87,chai98,Oxt90,sin91}, which uses a variational
free energy, has led to a good understanding of such freezing. It is
natural to ask whether there are nonequilibrium analogues of this
transition. One example is shear-induced melting of colloidal
crystals~\cite{das02}. We investigate another example of a
nonequilibrium transition in which dynamically generated turbulence
plays the role of temperature and disorders a crystal consisting of an
array of vortices, of alternating sign, imposed on a thin fluid film
by an external force.
 
Recent experiments~\cite{oue07,oue08} have explored this transition
from a low-Reynolds-number nonequilibrium vortex crystal, imposed on a
thin fluid film by a force that is periodic in space, to a
high-Reynolds-number, disordered, nonequilibrium liquid-type phase.
This problem has been studied earlier by using linear-stability
analysis~\cite{got84,got87} and direct numerical simulations
(DNS)~\cite{bra97,mol07}.  The former is well-suited to the study of
the first instability of the vortex crystal with increasing Reynolds
number $Re$; the latter have (a) studied the route to chaos, by using
techniques from nonlinear dynamics~\cite{bra97,mol07} to analyse the
temporal evolution of this system, or (b) have mimicked~\cite{bro08}
the experiments of Ref.~\cite{oue07} to study the creation and
annihilation of vortices.

We revisit the problem of turbulence-induced melting of a vortex
crystal by conducting a direct numerical simulation (DNS). Our study
combines methods from turbulence, nonlinear dynamics, and statistical
physics to elucidate the nature of nonequilibrium phases and
transitions in this system.  We use the two-dimensional (2D)
Navier-Stokes (NS) equation with air-drag-induced Ekman friction to
model the thin fluid films used in
experiments~\cite{oue07,oue08,riv01}; this is a good model for flows
in such films so long as the Mach number is small and the corrections
arising from the finite thickness of the film and from the Marangoni
effect can be neglected~\cite{cho90,cho01,per09}.  We force this 2D NS
equation in a manner that mimics the forcing used in experiments and
which yields, at low $Re$, a stationary, periodic array of vortices of
alternating signs; we refer to this array as the vortex crystal.

We then investigate the stability of this crystal as we increase the
amplitude of the force and, therefore, the Reynolds number.  Measures
from nonlinear dynamics, including time-series, power spectra, and
\Poincare-type maps, can be used fruitfully here to examine the
temporal behaviour of this system as it undergoes a sequence of
transitions.  This part of our study complements the work of
Refs.~\cite{bra97,mol07}.

Furthermore, we elucidate the natures of the transitions from the
spatially ordered crystal to the disordered, turbulent state by
borrowing ideas from the density-functional theory of the freezing of
a liquid into a crystal~\cite{ram79,chai98,Oxt90,sin91}.  Since the
density field of a crystal is {\it periodic}, this theory uses the
coefficients in the Fourier decomposition of the density as {\it order
  parameters}.  Specifically, in a conventional crystal, $\rho$ admits
the Fourier decomposition
\begin{equation}
  \rho({\bf r}) = \sum_{\bf G} 
  \rho_{\bf G} \exp(\imath {\bf G}
  \cdot{\bf r}) ,
\end{equation} 
where the sum is over the vectors {\bf G} of the {\it reciprocal
  lattice}; in the density-functional theory of
freezing~\cite{ram79,chai98} the Fourier coefficients $\rho_{\bf G}$
are identified as the order parameters of the liquid-to-crystal
transition since their mean values vanish in the liquid phase for all
nonzero reciprocal lattice vectors ${\bf G}$. Moreover, $\rho_{{\bf G}
  = 0}$ is the mean density and, because most liquids are nearly
incompressible, it undergoes a very small change at this transition.
Spatial correlations can be characterized conveniently by the
autocorrelation function $g({\bf r}) = \langle \overline{[\rho({\bf x}
  + {\bf r}) \rho({\bf x})]} \rangle$, where the angular brackets
denote Gibbsian thermal averages and the overline denotes spatial
averaging over ${\bf x}$; the Fourier transform of $g({\bf r})$ is
related~\cite{ram79,chai98} to the static structure factor $S({\bf
  k})$.  In an isotropic liquid phase the arguments of $g$ and $S$
are, respectively, $r = \mid {\bf r} \mid$ and $k = \mid {\bf k}
\mid$.

We use the Okubo-Weiss field $\Lambda\equiv \det(A)$ as the analogue
of the density $\rho({\bf r})$ in the density-functional theory of
freezing; here $A$ is the velocity-derivative matrix that has
components $A_{ij} \equiv \partial_i u_j$, with $u_j$ the $j^{th}$
component of the velocity.  For an inviscid, incompressible 2D fluid
the local flow topology can be characterized via the Okubo-Weiss
criterion~\cite{oku70,wei92}; this provides a useful measure of flow
properties even if viscosity and Ekman friction are
present~\cite{riv01,per09,pro93}: Regions with $\Lambda > 0$ and
$\Lambda < 0$ correspond, respectively, to centres and
saddles~\cite{oku70,wei92}.  Thus, in the nonequilibrium vortex
crystal, $\Lambda({\bf r})$ is a periodic function; so, like
$\rho({\bf r})$ in a conventional crystal, it admits the Fourier
decomposition
\begin{equation}
  \Lambda({\bf r}) = \sum_{\bf k} 
  {\hat{\Lambda}}_{\bf k} 
  \exp(\imath {\bf k} \cdot {\bf r}) ,
\end{equation}
where the sum is over the reciprocal-lattice vectors ${\bf k}$; it is
natural to think of ${\hat{\Lambda}}_{\bf k}$ as the order parameters
that characterise the vortex crystal. In terms of these order
parameters we can define the analogue of the static structure factor
$S({\bf k})$ for a conventional crystal; for the vortex crystal this
is the two-dimensional spectrum
\begin{equation}
  E_{\Lambda}({\bf k})\equiv 
  \langle {\hat{\Lambda}}_{\bf k}
  {\hat{\Lambda}}_{\bf -k} \rangle;
\end{equation}
here the angular brackets do not imply a Gibbsian thermal average, as
in equilibrium melting, but denote an average over the nonequilibrium
state of our system; as we show below, this state can be steady in
time, or it can oscillate periodically or quasiperiodically, or it
can be a chaotic state that is statistically steady.  The
autocorrelation function
\begin{equation}
  G({\bf r}) = \langle 
  \overline{\Lambda({\bf x + r}) 
    \Lambda({\bf x})} \rangle ,
\end{equation} 
is related to $E_{\Lambda}({\bf k})$ by a spatial Fourier transform.
The turbulent phase is nearly isotropic so, to a good approximation,
$G$ depends only on $r \equiv \mid {\bf r} \mid$ in this phase; and it
characterises the short-range order in the system exactly as $g(r)$
does in an isotropic liquid.

In addition to identifying a natural set of order parameters for the
turbulence-induced melting of a two-dimensional vortex crystal in a
thin fluid film, our study yields several interesting results that we
describe qualitatively below: In particular, we find that, as we
increase the Reynolds number, a rich sequence of transitions leads
from the steady ordered vortex crystal to the disordered, turbulent
state.  The third column in Table~\ref{tablech5:para} shows the types
of nonequilibrium phases we encounter: There is SX, the original, {\it
  steady} square crystal imposed by the force; this is followed by
SXA, {\it steady} crystals that are distorted, via large-scale spatial
undulations, relative to SX; these give way to distorted crystals that
oscillate in time, either periodically (OPXA) or quasiperiodically
(OQPXA); finally the system becomes disordered and displays {\it
  spatiotemporal chaos and turbulence} (SCT). OPXA and OQPXA are
actually a collection of several (perhaps an infinity) of
nonequilibrium crystalline phases that oscillate in time; given the
resolution of our calculation we can identify only some of these, as
we describe in detail below. The combination of techniques that we use
helps us to elucidate the natures of these phases in far greater
detail than has been attempted hitherto. We believe that, by using the
ideas we develop here, it should be possible to extend recent
experiments~\cite{oue07} on turbulence-induced melting to uncover the
rich series of nonequilibrium phase transitions that we have
summarised in Table 1.

The remaining part of this paper is organised as follows. In Sec. 2 we
present the equations we use to model turbulence-induced melting in
thin fluid films and the numerical methods we use. Section 3 contains
a detailed description of our results.  Section 4 is devoted to a
discussion and to conclusions.

\section{Model and Numerical Methods}

We begin with the 2D Navier-Stokes (NS) equations that can be written,
as follows, in the nondimensional form of Ref.~\cite{pla91}:
\begin{eqnarray}
  (\partial_t + {\bf u}\cdot \nabla) {\omega} 
  &=& \nabla^2 {\omega}/\Omega + 
  F_{\omega} - \alpha\omega;\label{eqch5:ns_vor}\\ 
  \nabla^2 \psi &=& \omega.
  \label{eqch5:poisson}
\end{eqnarray} 
Here ${\bf u}\equiv(-\partial_y \psi, \partial_x \psi)$, $\psi$, and
$\omega\equiv \nabla \times {\bf u}$ are, respectively, the velocity,
stream function, and vorticity at the position ${\bf x}$ and time $t$;
we choose the uniform density $\rho=1$; $\alpha$ is the
non-dimensionalised Ekman-friction coefficient, $\nu$ is the kinematic
viscosity, and $F_{\omega} \equiv -n^3 [\cos(n x) + \cos(n
y)]/\Omega$, is the non-dimensionalised force with injection wave
vector $n$, $\Omega=nF_{amp}/(\nu^2 k^3)$, and
$\alpha=n\nu\alpha^\prime k/F_{amp}$, where $F_{amp}$ is the forcing
amplitude, $\alpha^\prime$ is the Ekman friction, and lengths are
non-dimensionalised via a factor $k/n$ with $k$ a wave vector or
inverse length~\cite{pla91}. We denote the $x$ and $y$ components of
the velocity as $u_1\equiv u$ and $u_2 \equiv v$, respectively. The
spatially periodic force $F_{\omega}$ yields, at low $\Omega$, a
vortex crystal that is also referred to as a cellular flow.  A
linear-stability analysis of this flow indicates that it has a primary
instability~\cite{got87} at a critical Reynolds number
$Re_c\equiv\sqrt{2}$ which translates into a threshold value
$\Omega_{s,n} \equiv n Re_c$. This primary instability yields another
vortex crystal, which is steady in time but whose unit cell is larger
than that of the original vortex crystal~\cite{got87,the92b}.

We solve Eqs.~(\ref{eqch5:ns_vor}) and (\ref{eqch5:poisson})
numerically by using a pseudo-spectral method with a $2/3$ dealiasing
cut-off and a second-order Runge-Kutta scheme for time
marching~\cite{Can88,prasadthesis} with a time step $\delta t = 0.01$.
We use $N^2$ collocation points; in most of our studies we use
$N=128$; we have checked in representative cases that our results are
unchanged if we use $N=256$. Our main goal has been to obtain long
time series for several variables (see below) to make sure that the
temporal evolution of our system is obtained accurately; most of our
runs are at least as long as $3\times10^6 \delta t$. We monitor the
time-evolution of (a) the total kinetic energy $E(t)\equiv
\overline{{\bf u}^2}$, (b) the stream function $\psi$, (c) the
vorticity $\omega$, (d) the Okubo-Weiss parameter $\Lambda$, and (e)
the ${\bf k} = (1,0)$ component of the Fourier transform $\hat{v}$ of
the $y$ component $v$ of ${\bf u}$. Given these time series we obtain
$E_{\Lambda}({\bf k})$ at representative times and $G({\bf r})$, which
is obtained by averaging over $20$ configurations of $\Lambda({\bf
  r})$ separated from each other by $10^5 \delta t$, after transients
in the first $10^6$ time steps have been removed. From the time series
of $E(t)$ we obtain its temporal Fourier transform $E(f)$ and thence
the spectrum $\mid E(f) \mid$ that helps us to distinguish between
periodic, quasiperiodic, and chaotic temporal behaviours. We also
augment this charaterisation by using \Poincare-type sections in which
we plot $\Im\hat{v}_{(1,0)}$ versus $\Re\hat{v}_{(1,0)}$ at successive
times (see, e.g., Ref.~\cite{pla91} for the Kolmogorov flow).

We show below that the vortex crystal melts, as we increase $\Omega$,
via a rich sequence of transitions. The principal effect of the Ekman
friction is to delay the onsets of these transitions; we have checked
this explicitly in some cases.  However, to make contact with earlier
linear-stability and DNS studies of this problem, the results we
present below have been obtained with no Ekman friction. Our
qualitative conclusions are not affected by this.

So long as $\Omega < \Omega_{s,n}$, the steady-state
solution~\cite{pla91}, indicated by the subscript $s$, of
Eq.~(\ref{eqch5:ns_vor}) is $\omega_{s,n}=-n[\cos(nx) + \cos(ny)]$.
We examine the destabilisation of this state, with increasing
$\Omega$, for two representative values of $n$, namely, $n=4$ and
$n=10$ for which $\Omega_{s,4} \simeq 5.657$ and $\Omega_{s,10}
\simeq 14.142$; in our DNS we choose the initial vorticity field to have the
form $\omega=\omega_{s,n} + 10^{-4}\sum_{m_1=0,m_2=0}^{2,2}[\sin(m_1x
+ m_2y) + \cos(m_1x + m_2y)] m_2^2/\sqrt{(m_1^2+m_2^2)}$; and then we
let the system evolve under the application of the force $F_{\omega}$. 
We increase $\Omega$ from $\Omega_{s,n}$ to $6.85\Omega_{s,n}$ in
steps of $0.15\Omega_{s,n}$, for $n=4$ (runs {\tt R4-1} to {\tt R4-7} in
Table~\ref{tablech5:para}), and from $\Omega_{s,n}$ to
$15.9\Omega_{s,n}$ in steps of $0.1\Omega_{s,n}$, for $n=10$ 
(runs {\tt R10-1} to {\tt R10-4} in Table~\ref{tablech5:para}). 
To trace the transition to chaos accurately from SXA to SCT we have also 
conducted runs where $\Omega$ is increased in steps of $\simeq 0.08\Omega_{s,n}$
 for $n=4$ (runs {\tt R4-4} to {\tt R4-6} in Table~\ref{tablech5:para}) and 
$\simeq 0.07\Omega_{s,n}$ for $n=10$ (runs {\tt R10-2} and {\tt R10-3} in 
Table~\ref{tablech5:para}). We have benchmarked our numerical scheme by 
comparing results from our code with those of Ref.~\cite{pla91}, which deals 
with a Kolmogorov flow imposed by an external force of the form 
$F_{\omega} = n \cos(n y)$.

\begin{table}
  \begin{center}
    \begin{tabular}{@{\extracolsep{\fill}} c c c c c}
      \hline 
      $ Run $ & $n$ & $\Omega$ & $\rm{Order}$ \\ 
      \hline \hline 
      ${\tt R4-1}$ & $4$ & $\Omega <
      \Omega_{s,n}$ & $\rm{SX}$ \\
      ${\tt R4-2}$ & $4$ & $\Omega_{s,n} <
      \Omega \leq 6.5 $  & $\rm{SXA}$ \\
      ${\tt R4-3}$ & $4$ & $\Omega=8.202$
      & $\rm{OPXA}$ \\ 
      ${\tt R4-4}$ & $4$ &
      $9.05<\Omega<15.3$  & $\rm{SXA}$ \\
      ${\tt R4-5}$ &$4$ &
      $15.3<\Omega<17.3$ & $\rm{OPXA}$ \\ 
      ${\tt R4-6}$ & $4$ &$\Omega = 17.8$
      & $\rm{OQPXA + SCT}$\\ 
      ${\tt R4-7}$ & $4$ & $\Omega\geq
      18.3$ & $\rm{SCT}$ \\ 
      \hline \hline 
      ${\tt R10-1}$ & $10$ & $\Omega <
      \Omega_{s,n}$ & $\rm{SX}$ \\
      ${\tt R10-2}$ & $10$ & $\Omega_{s,n}
      <\Omega<22.6$ & $\rm{SXA}$ \\
      ${\tt R10-3}$ & $10$ &
      $24<\Omega<28$ & $\rm{OPXA}$ \\
      ${\tt R10-4}$ & $10$ &$\Omega\geq
      29$ & $\rm{SCT}$ \\ \hline
    \end{tabular}
  \end{center}
  \caption{Table indicating the number
    of the Run (e.g., R4-1), the values of 
    $n$ and $\Omega$, and the type of
    order. Here SX denotes the original,
    {\it steady} square crystal imposed by the
    force (the precise pattern depends
    on $n$); SXA denotes {\it steady} crystals
    that are distorted, via a
    large-scale undulation, relative to
    SX; OPXA indicates a crystal that is
    distorted slightly with respect to
    SX and which {\it oscillates
      periodically} in time; OQPXA is like
    OPXA but with {\it quasiperiodic
      oscillations}; SCT denotes the
    disordered phase with {\it spatiotemporal
      chaos and turbulence}.}
  \label{tablech5:para}
\end{table} 

\begin{figure*}[!p]
  \includegraphics[width=0.8\linewidth]{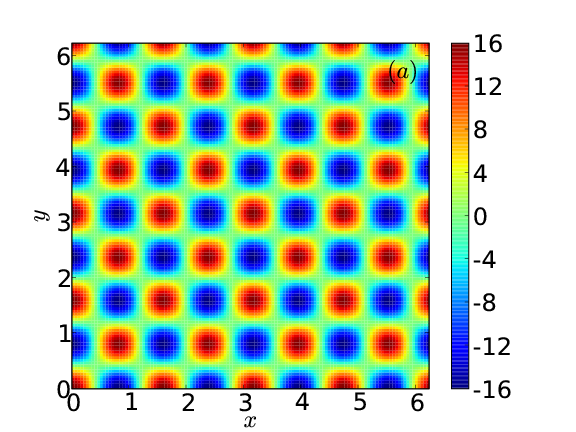}
  \includegraphics[width=0.8\linewidth]{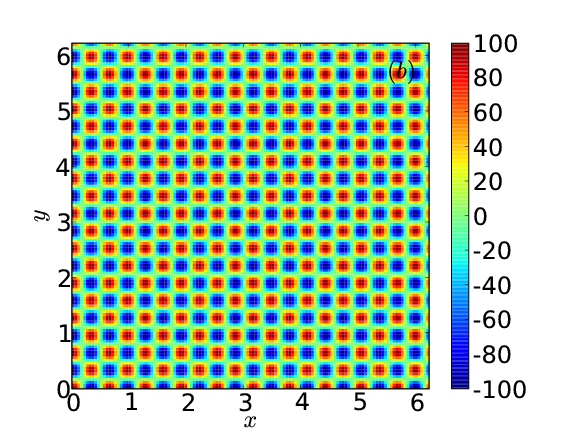}
  \caption{(Colour online) Pseudocolour plots, illustrating the vortex crystal 
for $\Omega<\Omega_{s,n}$, of the Okubo-Weiss field $\Lambda$ for
    (a)$n=4$, and (b) $n=10$. Given our colour bar, vortical regions,
    i.e., centres, appear red whereas strain-dominated regions, i.e.,
    saddles, appear dark blue.}
  \label{figch5:inlam}
\end{figure*}

\begin{figure*}[!p]
  \includegraphics[width=0.8\linewidth]{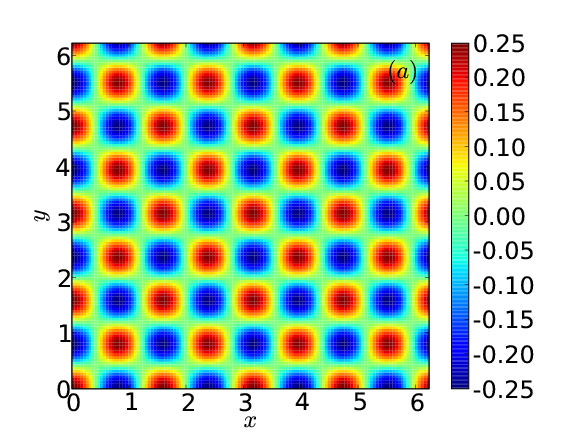}
  \includegraphics[width=0.8\linewidth]{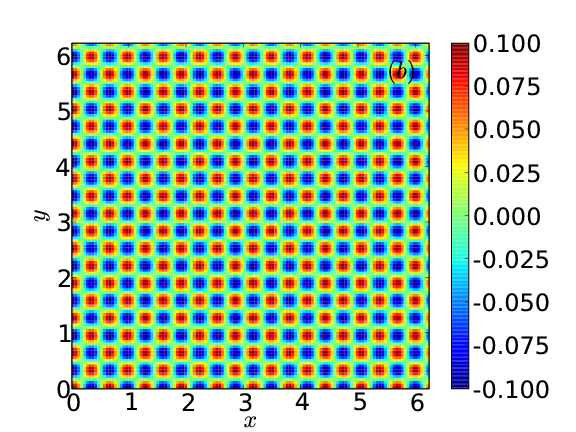}
  \caption{(Colour online) Pseudocolour plots, illustrating the vortex crystal 
for  $\Omega<\Omega_{s,n}$, of the streamfunction field $\psi$ for
    (a)$n=4$, and (b) $n=10$.}
  \label{figch5:inpsi}
\end{figure*}

For $\Omega < \Omega_{s,n}$, the $\Lambda$ field shows alternating
centres and saddles, arranged in a two-dimensional square lattice,
which we illustrate via pseudocolour plots for $n=4$ and $n=10$ in
Figs.~\ref{figch5:inlam} (a) and (b), respectively. [These patterns
are reminiscent of a two-dimensional version of a perfectly ordered
binary alloy, with two kinds of atoms, whose analogues here are
centres and saddles.] We show corresponding pseudocolour plots of
$\psi$ in Figs.~\ref{figch5:inpsi} (a) and (b), respectively.

\section{Results}

In this section we present the results of our numerical simulations
for $n=4$ and $n=10$ and the ranges of parameters given in
Table~\ref{tablech5:para}. In the first subsection we present our
results for $n=4$; the next subsection contains our results for
$n=10$; in the last subsection we describe our results for the order 
parameters and spatial correlation functions.

\subsection{The case $n=4$}

\begin{figure*}[!p]
  \begin{center}
    \includegraphics[width=0.5\linewidth]{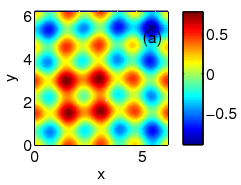}\\
    \includegraphics[width=0.5\linewidth]{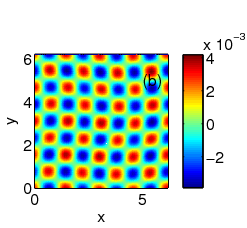}\\
    \includegraphics[width=0.5\linewidth]{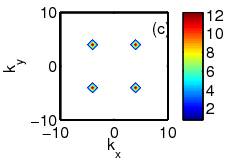}
  \end{center}
  \caption{\label{figch5:R42psilamk}(Colour online) Pseudocolour plots
    for $\Omega=6.5$ of (a) the streamfunction $\psi$ and (b) the
    Okubo-Weiss parameter $\Lambda$ with superimposed contour
    lines. (c) A filled contour plot of the reciprocal-space energy
    spectrum $E_{\Lambda}$ showing clear, dominant peaks at the
    forcing wave vectors.  }
\end{figure*}

When we increase $\Omega$ beyond $\Omega_{s,n=4}$, the steady-state
solution $\omega_{s,n=4}$ becomes unstable. From the range of values
of $\Omega$ in our runs ${\tt R4-2}$ (Table~\ref{tablech5:para}) we
observe that a new steady state is attained, which we illustrate, for
$\Omega = 6.5$, via pseudocolour plots of $\psi$ and $\Lambda$ in
Figs.~\ref{figch5:R42psilamk}(a) and (b), respectively.  The new
steady state is also a vortex crystal; however, it is different from
the original vortex crystal as can be seen especially clearly by
comparing the pseudocolour plots of $\psi$ in Figs.~\ref{figch5:inpsi}
(a) and \ref{figch5:R42psilamk}(a). This difference also shows up as a
very slight distortion of the crystalline structure in the
pseudocolour plot of $\Lambda$ shown in Fig.~\ref{figch5:R42psilamk}(b).
To use the language of solid state physics, this is an example of a
very weak {\it structural phase transition}. Normally such a phase
transition is mirrored in new {\it superlattice peaks} that appear in
the reciprocal-space spectrum $E_{\Lambda}$ in addition to the
dominant peaks associated with the original crystal structure;
however, given the weakness of the distortion, such superlattice peaks
are not visible, with our resolution in
Fig.~\ref{figch5:R42psilamk}(c).  Clear examples of such superlattice
peaks appear as we increase $\Omega$ (see below).

\begin{figure*}
  \includegraphics[width=0.45\linewidth]{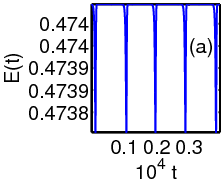}
  \includegraphics[width=0.45\linewidth]{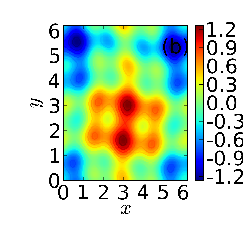}\\
  \includegraphics[width=0.45\linewidth]{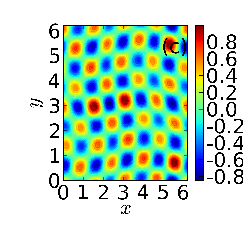}
  \includegraphics[width=0.45\linewidth]{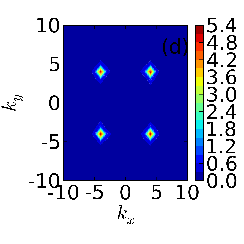}
  \caption{\label{figch5:R43psilamk}(Colour online) Plot for
    $\Omega=8.202$ of (a) the time evolution of the energy $E(t)$ for
    $\Omega=8.202$.  Pseudocolour plots of (b) the streamfunction
    $\psi$ and (c) the Okubo-Weiss parameter $\Lambda$ with
    superimposed contour lines.  (d) A filled contour plot of the
    reciprocal-space energy spectrum $E_{\Lambda}$ showing the
    distortions of peaks at the forcing wave vectors.  }
\end{figure*}

At $\Omega=8.202$, a new regime appears (run ${\tt R4-3}$). The
time-series of the energy $E(t)$ now shows a periodic array of spikes.
This regime has no analogue in a conventional crystal; it is a
crystal that oscillates periodically in time and, to that extent, it
can be thought of as a {\it spatiotemporal crystal}. The time between
successive spikes is very large ($\simeq 10^4 \delta t$) as shown by
the plot of $E(t)$ in Fig.~\ref{figch5:R43psilamk}(a); this is why our
DNS runs must be very long to distinguish such states from one that is
steady; we have also checked that the time between successive spikes
is the same (to three-figure accuracy) for $N=64$ and $N=128$. In
Figs.~\ref{figch5:R43psilamk}(b) and (c) we show pseudocolour plots of
$\psi$ and $\Lambda$, respectively; the former shows a large-scale
spatial undulation and the latter some deformation relative to the
original crystal. This deformation is also mirrored in the distortion,
relative to Fig.~\ref{figch5:R42psilamk}(c), of the dominant peaks in
the reciprocal-space spectrum $E_{\Lambda}({\bf k})$ shown in
Fig.~\ref{figch5:R43psilamk}(d).

\begin{figure*}[!p]
  \begin{center}
    \includegraphics[width=0.45\linewidth]{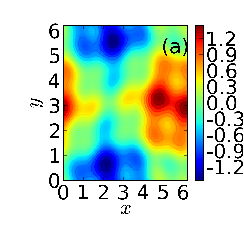}\\
    \includegraphics[width=0.45\linewidth]{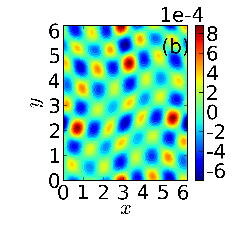}\\
    \includegraphics[width=0.45\linewidth]{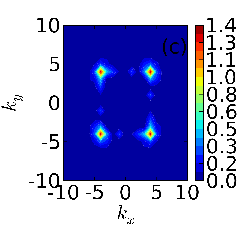}
  \end{center}
  \caption{(Colour online) Pseudocolour plots for $\Omega=11.3$ of (a)
    the streamfunction $\psi$ and (b) the Okubo-Weiss parameter
    $\Lambda$ with superimposed contour lines. (c) A filled contour
    plot of the reciprocal-space energy spectrum $E_{\Lambda}$ showing
    clear, dominant peaks, at the forcing wave vectors, and
    subdominant superlattice peaks at smaller wave vectors.  }
  \label{figch5:R44psilamk}
\end{figure*}

For runs ${\tt R4-4}$, i.e., $9.05 \leq \Omega < 15.3$, we find a new
crystalline state that is steady in time. It has a large-scale spatial
undulation relative to the original vortex crystal as illustrated, for
$\Omega = 11.3$, by the pseudocolour plots of $\psi$ and $\Lambda$ in
Figs.~\ref{figch5:R44psilamk}(a) and (b), respectively. This
undulation leads to a distortion of the dominant peaks in the
reciprocal-space spectrum $E_{\Lambda}({\bf k})$ of
Fig.~\ref{figch5:R44psilamk}(c), which also shows new superlattice
peaks that occur at smaller values of $(k_x,k_y)$ relative to the
dominant peaks.

\begin{figure*}[!p]
  \begin{center}
    \includegraphics[width=0.45\linewidth]{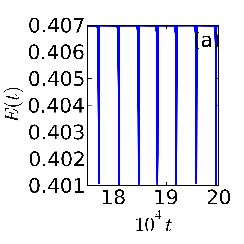}\\
    \includegraphics[width=0.45\linewidth]{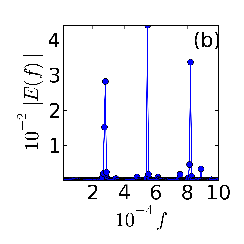}\\
    \includegraphics[width=0.45\linewidth]{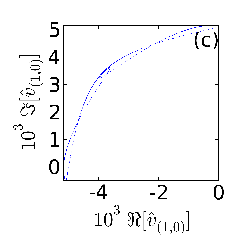}
  \end{center}
  \caption{\label{figch5:R45tsf}(Colour online) Plots for
    $\Omega=15.3$ of (a) the time evolution of the energy $E(t)$, (b)
    $\mid E(f) \mid$ versus the frequency $f$, and (c) the
    \Poincare-type section in the $(\Re\hat{v}[1,0],\Im\hat{v}[1,0])$
    plane.  }
\end{figure*}

\begin{figure*}[!p]
  \includegraphics[width=0.48\linewidth]{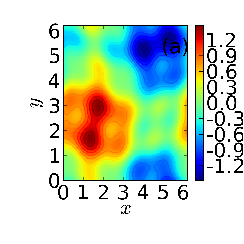}
  \includegraphics[width=0.48\linewidth]{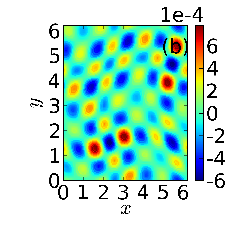}\\
  \includegraphics[width=0.48\linewidth]{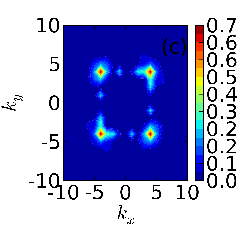}
  \includegraphics[width=0.48\linewidth]{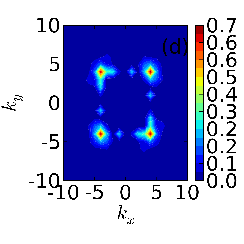}
  \caption{\label{figch5:R45psilamk}(Colour online) Pseudocolour plots
    for $\Omega=15.3$ of (a) the streamfunction $\psi$ and (b) the
    Okubo-Weiss parameter $\Lambda$ with superimposed contour lines.
    (c) and (d) Filled contour plots of $E_{\Lambda}$ showing the two
    spectra between which it oscillates in time.  }
\end{figure*}

On further increasing $\Omega$ we enter a new regime (runs ${\tt
  R4-5}$, i.e., $15.3 \leq \Omega < 17.3$) in which we have a
spatiotemporal crystal, i.e., a spatially periodic $\Lambda$ that
oscillates in time. The time-series $E(t)$ displays a periodic array
of spikes as shown in Fig.~\ref{figch5:R45tsf}(a); this leads to the
frequency-space spectrum $\mid E(f)\mid$ of
Fig.~\ref{figch5:R45tsf}(b). The peaks in this spectrum can be
labelled as $\ell f_0$, where $\ell$ is a positive integer and $f_0$
is the fundamental frequency that can be obtained from the inverse of
the temporal separation between successive spikes in
Fig.~\ref{figch5:R45tsf}(a); this is a clear signature of periodic
temporal evolution. The \Poincare-type map in the $(\Re[\hat{v}(1,0)],
\Im[\hat{v}(1,0)])$ plane, Fig.~\ref{figch5:R45tsf}(c), shows that the
projection of the attractor on this plane is a closed loop in this
case. Pseudocolour plots of $\psi$ and $\Lambda$ [
Figs.~\ref{figch5:R45psilamk}(a)-(b)] are similar, respectively, to
those in Figs.~\ref{figch5:R44psilamk}(a)-(b) if we look at their
spatial patterns; however, they oscillate in time as can be seen most
clearly from their animated versions [mpeg files
psi\_movie\_R5.mpeg-lam\_movie\_R5.mpeg].  The associated
reciprocal-space spectrum $E_{\Lambda}$ also oscillates between the
spectra shown in Figs.~\ref{figch5:R45psilamk}(c) and (d) as can be
seen clearly from its animated version [avi file lamf\_movie\_R5.avi].

\begin{figure*}[!p]
  \begin{center}
    \includegraphics[width=0.6\linewidth]{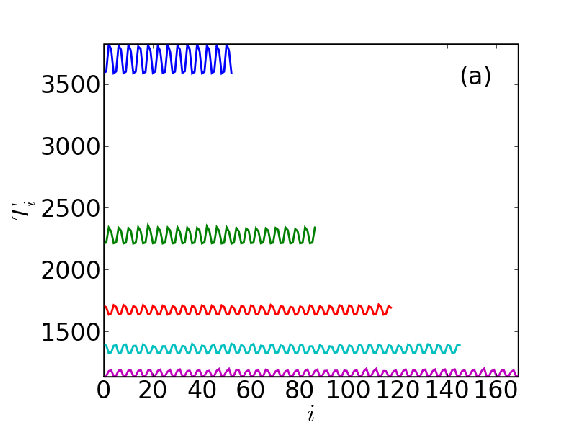}\\
    \includegraphics[width=0.6\linewidth]{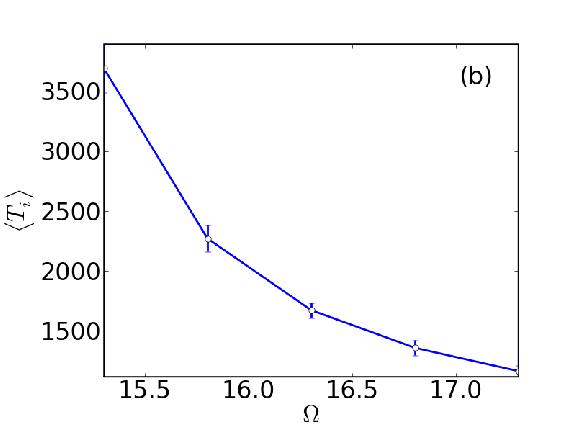}
  \end{center}
  \caption{\label{figch5:interval4}(Colour online) Plots of (a) the
    inter-spike interval $T_i$ versus the spike index $i$ for
    $\Omega=15.3$ (red curve), $\Omega=15.8$ (black curve),
    $\Omega=16.3$ (purple curve), $\Omega=16.8$ (green curve), and
    $\Omega=17.3$ (blue curve) and (b) the time mean value of $T_i$
    versus $\Omega$ (see text for error bars).}
\end{figure*}

The time between successive spikes in $E(t)$ decreases as $\Omega$
increases. To quantify this, we define the inter-spike interval $T_i$
as follows: $T_i \equiv t_{i+1}-t_i$, where $t_i$ is the time at which
$E(t)$ crosses, for the $i^{th}$ time, its mean value, $\langle
E(t)\rangle$, from below; we can think of $i$ as the spike index.  In
Fig.~\ref{figch5:interval4}(a) we plot $T_i$ versus $i$; this shows
that the mean value of $T_i$ decreases as $\Omega$ increases
[Fig.~\ref{figch5:interval4}(b)]; furthermore, $T_i$ oscillates
slightly about its mean value for any given value of $\Omega$. The
magnitude of these oscillations, which we have used for the error bars
in Fig.~\ref{figch5:interval4}(b), decreases as $\Omega$ increases. We
have checked explicitly that our results here do not change when we
increase the resolution of our DNS from $N=128$ to $N=256$.

\begin{figure*}
  \includegraphics[width=0.45\linewidth]{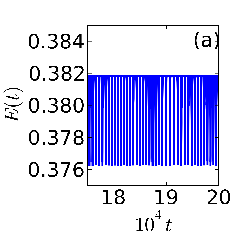}
  \includegraphics[width=0.45\linewidth]{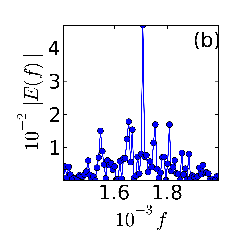}
  \includegraphics[width=0.45\linewidth]{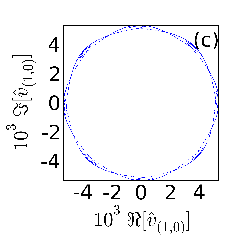}
  \includegraphics[width=0.45\linewidth]{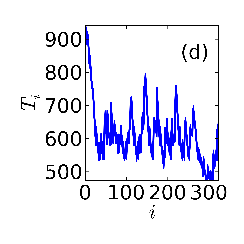}
  \caption{\label{figch5:R46tsf}(Colour online) Plots for
    $\Omega=17.8$ of (a) the time evolution of the energy $E(t)$, (b)
    the spectrum $E(f)$ versus $f$, (c) the \Poincare-type section in
    the plane $(\Re[\hat{v}(1,0)],\Im[\hat{v}(1,0)])$, and (d) the
    inter-spike interval $T_i$ versus the spike index $i$.  }
\end{figure*}

At $\Omega=17.8$, i.e., run ${\tt R4-6}$, another transition occurs:
The time series of the energy $E(t)$ and its frequency spectrum $\mid
E(f) \mid$ are shown, respectively, in Figs.~\ref{figch5:R46tsf}(a)
and (b). The latter displays peaks superimposed on a noisy background
signal; these peaks can be indexed as $f_0, f_1, f_1-2f_0, f_0-2f_1,$
and $3f_0-2f_1$, within our numerical accuracy and with $f_0
\simeq 0.001653$
and $f_1 \simeq 0.001707$.  Since $f_0/f_1$ is not a simple rational number,
we conclude that Fig.~\ref{figch5:R46tsf}(b) indicates principally
quasiperiodic temporal evolution with a small chaotic admixture, the
former associated with the peaks indexed above and the latter with the
noisy background signal. We believe the chaotic part of the signal
comes from transitions between the elliptical islands in the
\Poincare-type section of Fig.~\ref{figch5:R46tsf}(c).  The plot of
the inter-spike interval $T_i$ versus the spike index $i$ in
Fig.~\ref{figch5:R46tsf}(d) confirms the complicated temporal
evolution of this state.

\begin{figure*}[!p]
  \begin{center}
    \includegraphics[width=0.45\linewidth]{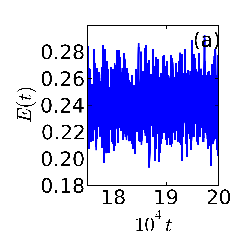}\\
    \includegraphics[width=0.45\linewidth]{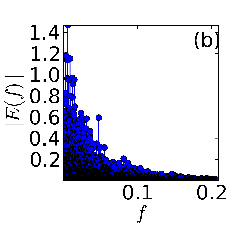}\\
    \includegraphics[width=0.45\linewidth]{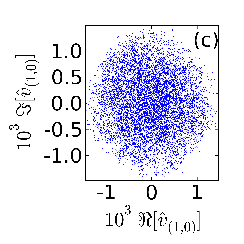}
  \end{center}
  \caption{(Colour online) Plots for $\Omega=50$ of (a) the time evolution 
of the energy $E(t)$, (b) the spectrum $\mid E(f)\mid$ versus $f$, and
    (c) the \Poincare-type section in the
    $(\Re[\hat{v}(1,0)],\Im[\hat{v}(1,0)])$ plane.  }
  \label{figch5:R47tsf}
\end{figure*}

\begin{figure*}[!p]
  \begin{center}
    \includegraphics[width=0.45\linewidth]{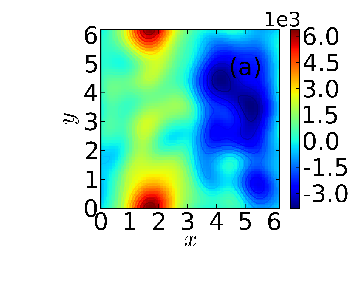}\\
    \includegraphics[width=0.45\linewidth]{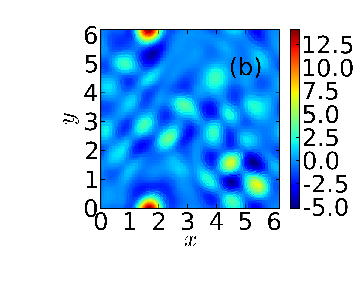}\\
    \includegraphics[width=0.45\linewidth]{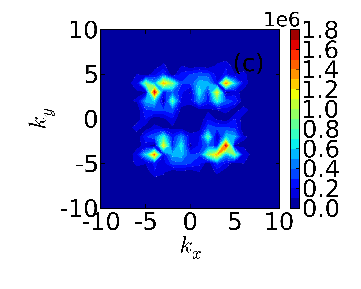}
  \end{center}
  \caption{(Colour online) Pseudocolour plots for $\Omega=50$ of (a)
    the streamfunction $\psi$ and (b) the Okubo-Weiss parameter
    $\Lambda$ with superimposed contour lines. (c) A filled contour
    plot of the reciprocal-space energy spectrum $E_{\Lambda}$ which
    shows that a large number of modes are excited.  }
  \label{figch5:R47psilamk}
\end{figure*}

As we increase $\Omega$ further (${\tt R4-7}$) the temporal evolution
of the system becomes ever more chaotic; this is associated with a
disordered pattern of vortices in space too.  Thus we obtain a state
with spatiotemporal chaos and turbulence, which is our analogue of the
liquid state. We illustrate this for $\Omega=50$.  We begin with the
time series of $E(t)$ and the spectrum $\mid E(f)\mid$ in
Figs.~\ref{figch5:R47tsf}(a) and (b), respectively; the latter shows
clearly a broad background that is indicative of temporal chaos. This
is further confirmed by the nearly uniform spread of points in the
\Poincare-type section [Fig.~\ref{figch5:R47tsf}(c)] in the
$(\Re[\hat{v}(1,0)],\Im[\hat{v}(1,0)])$ plane.  The disordered spatial
organisation of this state is illustrated by the pseudocolour plots of
$\psi$ and $\Lambda$ [Figs.~\ref{figch5:R47psilamk}(a)-(b),
respectively] and the reciprocal-space spectrum $E_\Lambda({\bf k})$
of Fig.~\ref{figch5:R47psilamk}(c) that shows several new modes in
addition to the original peaks, whose vestiges are still visible since
$F_{\omega}$ continues to act on the system.

Thus we see that the turbulence-induced melting of our nonequilibrium
vortex crystal is far richer than its equilibrium counterpart.  For
the case $n=4$ investigated above it proceeds as described in
column 4 in Table~\ref{tablech5:para}. Before we present an analysis
of the disordered state in terms of the spatial autocorrelation
function $G({\bf r})$ and the evolution of the order parameters
$\langle {\hat \Lambda_{\bf k}} \rangle$ with $\Omega$, we give below
a short summary of our results for $n = 10$; the route to turbulence
is slightly different for this case than for $n=4$.


\subsection{The case n=10}

Our results for $n=10$ are based on the runs ${\tt R10-1}$ to ${\tt
  R10-5}$ in Table~\ref{tablech5:para}.

\begin{figure*}
  \begin{center}
    \includegraphics[width=0.45\linewidth]{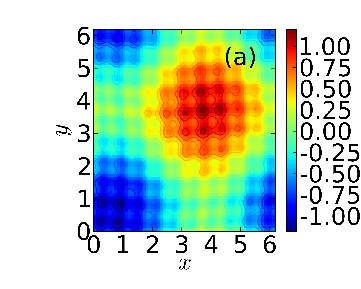}\\
    \includegraphics[width=0.45\linewidth]{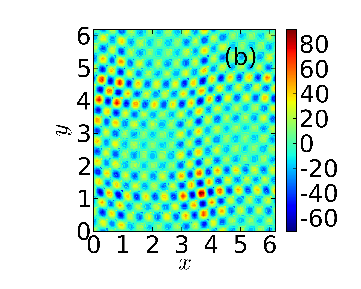}\\
    \includegraphics[width=0.45\linewidth]{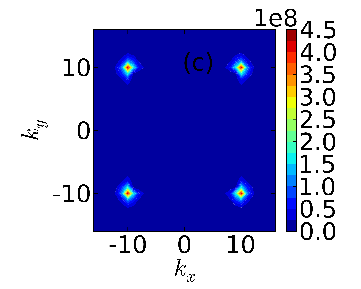}
  \end{center}
  \caption{\label{figch5:R102psilamk}(Colour online) Pseudocolour
    plots for $\Omega=22.62$ of (a) the streamfunction $\psi$ and (b)
    the Okubo-Weiss parameter $\Lambda$ with superimposed contour
    lines.  (c) A filled contour plot of the reciprocal-space energy
    spectrum $E_{\Lambda}$ showing clear, but slightly distorted,
    dominant peaks at the forcing wave vectors.  }
\end{figure*}

For $\Omega<\Omega_{s,n}$ the steady vortex crystal is shown by the
pseudocolour plot of $\Lambda$ in Fig.~\ref{figch5:inlam}(b).  As we
increase $\Omega$ beyond $\Omega_{s,n}$ we find in runs ${\tt R10-2}$,
i.e., the range $\Omega_{s,n} < \Omega < 22.6$, a new steady state in
which pseudocolour plots of $\psi$ and $\Lambda$ show large-scale
spatial undulations caused by small deformations of the original
vortex crystal [Figs.~\ref{figch5:R102psilamk} (a) and (b),
respectively]; consequently the dominant peaks in the reciprocal-space
spectrum $E_{\Lambda}$ are slightly distorted.

\begin{figure*}[!p]
  \begin{center}
    \includegraphics[width=0.45\linewidth]{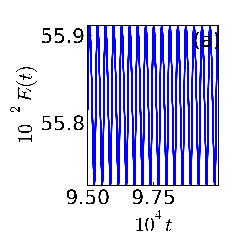}\\
    \includegraphics[width=0.45\linewidth]{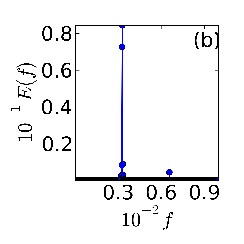}\\
    \includegraphics[width=0.45\linewidth]{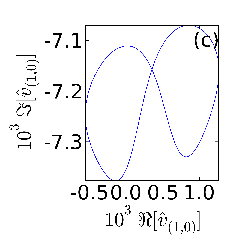}
  \end{center}
  \caption{\label{figch5:R103tsf}(Colour online) Plots for $\Omega=24$ of (a) the
    time evolution of energy $E(t)$ versus $t$, (b) the power spectrum
    $E(f)$ versus $f$, and (c) the \Poincare-type section
    $\Im[\hat{v}(1,0)]$ versus $\Re[\hat{v}(1,0)]$.  }
\end{figure*}

\begin{figure*}[!p]
  \begin{center}
    \includegraphics[width=0.45\linewidth]{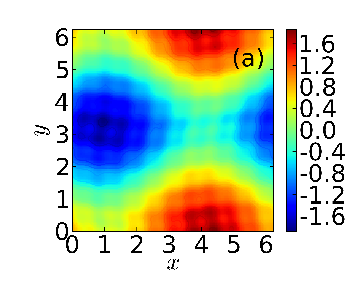}\\
    \includegraphics[width=0.45\linewidth]{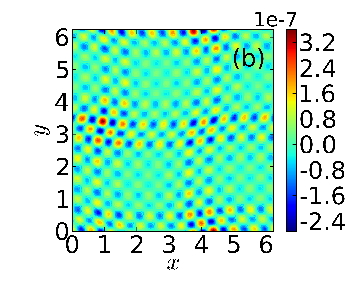}\\
    \includegraphics[width=0.45\linewidth]{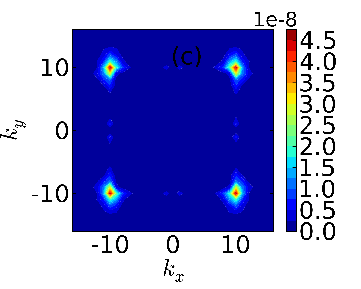}
  \end{center}
  \caption{\label{figch5:R103psilamk}(Colour online) Pseudocolour
    plots for $\Omega=24$ of (a) the streamfunction $\psi$ and (b) the
    Okubo-Weiss parameter $\Lambda$ with superimposed contour lines.
    (c) A filled contour plot of the reciprocal-space energy spectrum
    $E_{\Lambda}$ showing clear, dominant peaks at the forcing wave
    vectors.  }

\end{figure*}

Around $\Omega=24$ (run ${\tt R10-3}$) another transition occurs: the
time series of $E(t)$ is periodic [Fig.~\ref{figch5:R103tsf}(a)] and
its spectrum $\mid E(f) \mid$ [Fig.~\ref{figch5:R103tsf}(b)] shows one
dominant peak, i.e., higher harmonics are nearly absent. Thus the
\Poincare-type plot in the $(\Re\hat{v}[1,0],\Im\hat{v}[1,0])$ plane
[Fig.~\ref{figch5:R103tsf}(c)] displays a simple attractor.  The
spatial structure of this state is illustrated by the pseudocolour
plots of $\psi$ and $\Lambda$ shown, respectively, in
Figs.~\ref{figch5:R103psilamk}(a) and (b); the associated
reciprocal-space spectrum $E_{\Lambda}$ is shown in
Fig.~\ref{figch5:R103psilamk}(c).  Given the temporal behaviour of
this state, these structures, in real or reciprocal space, oscillate
in time at the frequency given by the temporal evolution of
$E(t)$. For $\Omega=24$ the large-scale spatial structures in $\psi$
oscillate around their mean positions; but, for $25 \leq \Omega \leq
28$, we find a travelling-wave type pattern, which reenters our
simulation domain by virtue of the periodic boundary conditions that
we use in our pseudo-spectral method.  If we compare the frequency
spectra $\mid E(f) \mid$ for the cases $\Omega = 24$ and $\Omega =
28$, we find higher harmonics in the latter but they are all multiples
of one fundamental frequency.


\begin{figure}[!p]
  \begin{center}
    \includegraphics[width=0.45\linewidth]{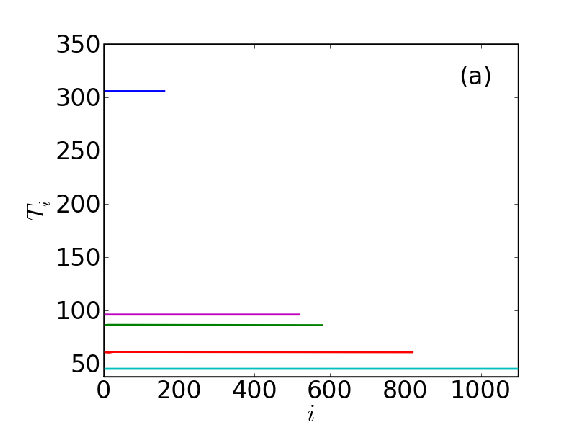}
    \includegraphics[width=0.45\linewidth]{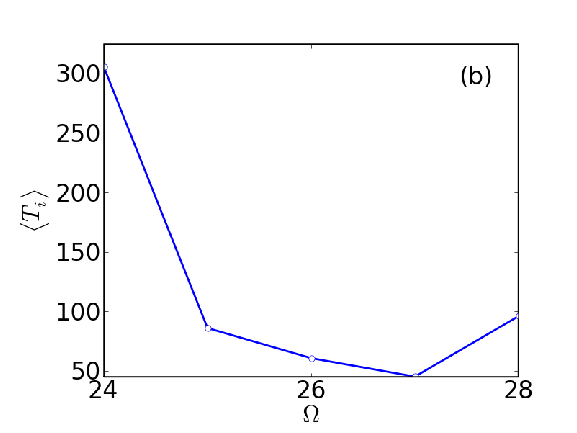}
  \end{center}
  \caption{\label{figch5:interval5}(Colour online) Plots of (a) the 
inter-spike interval $T_i$ versus the spike index $i$ for $\Omega=24$ 
(blue curve), $\Omega=25$ (purple curve), $\Omega=26$ (green curve), 
$\Omega=27$ (red curve), and $\Omega=28$ (cyan curve) and (b) the 
time mean value $T_i$ for different values of $\Omega$.}
\end{figure}

In Fig.~\ref{figch5:interval5}(a) we show plots of $T_i$ versus $i$
(cf. Fig.~\ref{figch5:interval4}(a) for $n=4$) for various values of
$\Omega$. From these we obtain the plot of the mean value $\langle T_i
\rangle$ versus $\Omega$ shown in Fig.~\ref{figch5:interval5}.  This
first decreases, as we increase $\Omega$, and then increases mildly at
$\Omega=28$.

\begin{figure*}[!p]
  \begin{center}
    \includegraphics[width=0.45\linewidth]{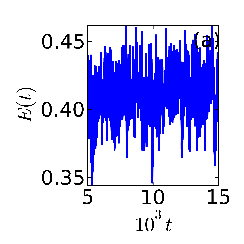}\\
    \includegraphics[width=0.45\linewidth]{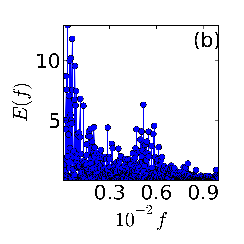}\\
    \includegraphics[width=0.45\linewidth]{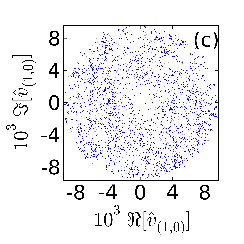}
  \end{center}
  \caption{\label{figch5:R104tsf} (Colour online) Plots for $\Omega=225$ of (a) the
    time evolution of $E(t)$ , (b) the spectrum $\mid E(f) \mid$
    versus $f$, and (c) the \Poincare-type section in the
    $(\Re[\hat{v}(1,0)],\Im[\hat{v}(1,0)])$ plane.  }
\end{figure*}

\begin{figure*}
  \begin{center}
    \includegraphics[width=0.45\linewidth]{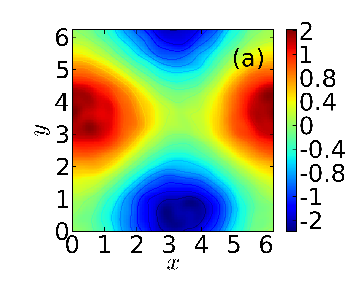}\\
    \includegraphics[width=0.45\linewidth]{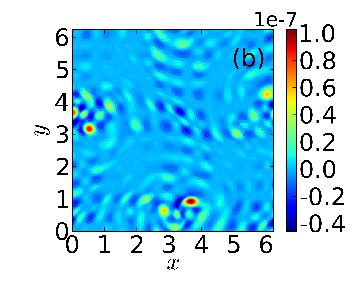}\\
    \includegraphics[width=0.46\linewidth]{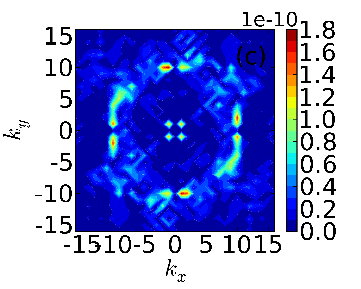}
  \end{center}
  \caption{\label{figch5:R104psilamk}(Colour online) Pseudocolour
    plots for $\Omega=225$ of (a) the streamfunction $\psi$ and (b)
    the Okubo-Weiss parameter $\Lambda$ with superimposed contour
    lines.  (c) A filled contour plot of the reciprocal-space energy
    spectrum $E_{\Lambda}$ showing that a large number of modes are
    excited.  }
\end{figure*}

For $\Omega \geq 29$ the time-series of $E(t)$ appears chaotic and the
associated frequency spectrum $\mid E(f) \mid$ displays a broad
background, as we show in the illustrative Figs.~\ref{figch5:R104tsf}
(a) and (b) for $\Omega = 225$. The associated \Poincare-type section
in Fig.~\ref{figch5:R104tsf}(c) confirms that the temporal behaviour
is chaotic. The spatial patterns are also disordered as we show via
the pseudocolour plots of $\psi$ and $\Lambda$ in
Figs.~\ref{figch5:R104psilamk}(a) and (b), respectively.  The
corresponding reciprocal-space spectrum $E_{\Lambda}$ shows that a
large number of modes are excited.  Thus we have both spatial disorder
and temporal chaos; as in the case $n=4$, the analogue of the liquid
state is a turbulent one with spatiotemporal chaos. However, given the
resolution in $\Omega$ that we have been able to obtain in our
calculations, the route to this state of spatiotemporal chaos is
different for $n=10$ and $n=4$ as can be seen from column 4 in
Table~\ref{tablech5:para}.


\subsection{Order parameters and spatial autocorrelation functions}

\begin{figure}[!p]
  \begin{center}
    \includegraphics[width=0.6\linewidth]{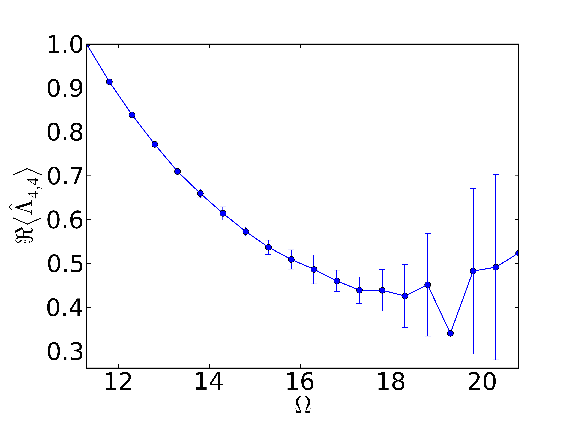}\\
    \includegraphics[width=0.6\linewidth]{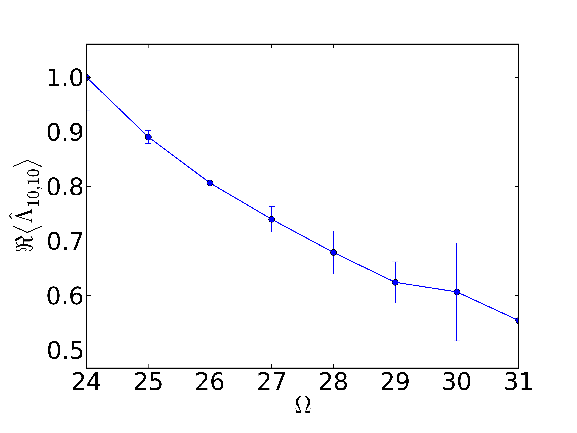}
  \end{center}
  \caption{\label{figch5:rhog}(Colour online) Plot showing a decrease in
    $\langle{\hat{\Lambda}}_{\bf k}\rangle$ with increasing $\Omega$
    for (a) $n=4$ and ${\bf k} = (4,4)$ and (b) $n=10$ and ${\bf k} =
    (10,10)$.}
\end{figure}

We now return to ideas borrowed from the density-functional
theory~\cite{ram79,chai98,Oxt90,sin91} of freezing by examining the
behaviour of the order parameters $\langle{\hat{\Lambda}}_{\bf
  k}\rangle$ as functions of $\Omega$.  Recall that, in equilibrium
melting, $\rho_{\bf G}$ jumps discontinuously from a nonzero value in
the crystal to zero in the liquid at the first-order melting
transition. As we have noted above, the turbulence-induced melting of
our vortex crystal is far more complicated; it proceeds via a sequence
of transitions. In turbulence-induced melting of a vortex crystal 
$\Re\langle{\hat{\Lambda}}_{\bf k}\rangle$, obtained by summing 
${\hat{\Lambda}}_{\bf k}$ over 
the four forcing wave vectors, is the equivalent of 
$\Re\langle\rho_{\bf G}\rangle$ 
in a conventional crystal; $\Re\langle{\hat{\Lambda}}_{\bf k}\rangle$ changes
with $\Omega$ as shown, respectively, for (a) $n=4$ and ${\bf k} =
(4,4)$ and (b) $n=10$ and ${\bf k} = (10,10)$ in 
Figs.~\ref{figch5:rhog} (a) and (b). For small values of $\Omega$  
the steady state is SX so, in Fourier space, only modes with the forcing 
wave vector are significant. On increasing $\Omega$,  
the crystal undergoes a series of transitions that take it from 
the crystal SX to the disordered, turbulent state SCT; as this
happens the spectral weight slowly shifts away the forcing wave
numbers. This explains the trend observed in Figs.~\ref{figch5:rhog}.

\begin{figure*}[!p]
  \includegraphics[width=0.45\linewidth]{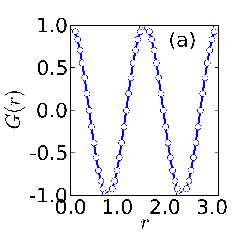}
  \includegraphics[width=0.45\linewidth]{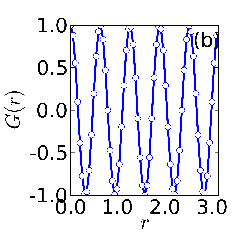}
  \includegraphics[width=0.45\linewidth]{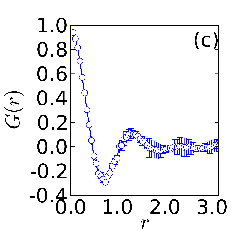}
  \includegraphics[width=0.45\linewidth]{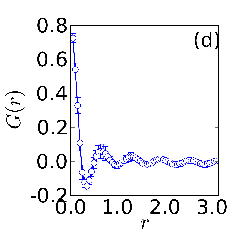}
  \caption{\label{figch5:tranmelt} (Colour online) Plots of
    $G({\bf r})$: (i) crystalline state: (a) $n=4, \Omega<\Omega_{s,n}$
    and (b) $n=10, \Omega<\Omega_{s,n}$ along the line connecting
    ${\bf r} =
    (\pi/2,\pi/2)$ and ${\bf r} = (\pi/2,\pi)$; and (ii)
    circularly averaged in the
    turbulent state: (c)
    $n=4$, $\Omega=20.81$ and (d) $n=10, \Omega=225$.}
\end{figure*}

The spatial correlations in the crystalline and disorderd states can be
characterised by the spatial autocorrelation function $G({\bf
  r})$ defined in Eq. (4). Representative plots are shown, respectively, for $n=4$ and
$n=10$ in Figs.~\ref{figch5:tranmelt} (a)-(d).  For the crystalline
case we evaluate $G({\bf r})$ along the line connecting ${\bf r} =
(\pi/2,\pi/2)$ and ${\bf r} = (\pi/2,\pi)$; this shows a periodic
array of peaks [Figs.~\ref{figch5:tranmelt} (a) and (b) for $n=4$ and
$n=10$, respectively]; the widths of these peaks are related to the
widths of vortical or strain-dominated regions.  In the turbulent
phase we present data obtained by a circular average of $G$
[Figs.~\ref{figch5:tranmelt} (c) and (d) for $n=4$ and $n=10$,
respectively]; here the peaks decay over a length scale that indicates
the degree of short-range order.  This decay is similar to the
decay of
spatial correlation functions in a disordered liquid.

\section{Conclusions}

We have carried out a detailed numerical study of turbulence-induced
melting of a nonequilibrium vortex crystal in a forced, thin fluid
film. We use ideas from the density-functional theory of
freezing~\cite{ram79,chai98,Oxt90,sin91}, nonlinear dynamics, and
turbulence to characterise this. Correlation functions, similar to
those in liquid-state theory, have been used by some recent
experiments~\cite{oue07,oue08} to analyse the short-range order in the
turbulent phase; nonlinear-dynamics methods, such as \Poincare-type
maps, have been used in the numerical studies of Ref.~\cite{bra97};
experimental studies have used the curvature of Lagrangian
trajectories to identify extrema in vortical and strain-dominated
regimes. To the best of our knowledge, there is no study that brings
together a variety of methods, as we do, to analyse turbulence-induced
melting.

The advantages of our approach are as follows: (a) it helps us to
identify the order parameters for turbulence-induced melting and thus
contrast it with conventional melting; (b) the sequence of transitions
can be characterised completely in terms of the Eulerian fields $\psi$
and $\Lambda$, the total energy $E(t)$, and suitable Fourier
transforms of these; (c) spatial correlations in crystalline and
turbulent phases can be studied conveniently in terms of $G$.

Equilibrium phase transitions occur strictly only in the thermodynamic
limit that is, roughly speaking, the limit of infinite
size~\cite{ruelle}. It is
interesting to ask how we might take the thermodynamic limit for
the vortex crystals we have studied here.  There seem to be at least two
ways to do this: (a) in the first the system size can be taken to
infinity in such a way that the areal density of the vortical and
strain-dominated regimes remains the same in the ordered, crystalline
phase; (b) in the second way we can increase the parameter $n$ in the forcing
$F_{\omega}$ so that more and more unit cells occupy the simulation
domain (see, e.g., Figs.~\ref{figch5:inlam}(a) and (b) for $n=4$ and
$n=10$, respectively).  Such issues have not been addressed in detail
by any study, partly because, for large system sizes, it is not
possible to obtain the long time series that are required to
characterise the temporal evolution of the system (especially in the
states we have referred to as spatiotemporal crystals). In particular,
it is quite challenging to investigate the system-size dependence of
the transitions summarised for $n=4$ and $n=10$ in
Table~\ref{tablech5:para}. From Figs. \ref{figch5:tranmelt} (c) and (d) 
we can extract a correlation length; this length is much smaller than 
the linear size of our simulation domain so we expect that our results 
in the turbulent phase (SCT) will not change if we increase the size of 
our system. However, subtle size dependence might occur in the ordered 
phase as follows: as we increase $\Omega$, the original crystal is 
distorted by large-scale spatial undulations that are associated with 
the inverse cascade of energy in two-dimensional turbulence; if these 
undulations lead to crystalline phases with larger and larger unit cells, 
then our finite-size calculations will become unreliable when the size 
of the unit cell becomes comparable to the size of our simulation domain. 
A systematic study of such subtle finite-size effects is a very challenging 
task that requires much more detailed simulations than have been attempted 
so far.

As we have shown above, the sequence of transitions that comprise
turbulence-induced melting of a vortex crystal is far richer than
conventional equilibrium melting.  There is another important way in
which the former differs from the latter: To maintain the steady
states, statistical or strictly steady, we always have
a force $F_{\omega}$; thus, in the language of phase transitions, we
always have a symmetry-breaking field, both in the ordered and
disordered phases. Strictly speaking, therefore, there is no symmetry
difference between the disordered, turbulent state and the
ordered vortex crystal, as can be seen directly from the remnants of the dominant
peaks in the reciprocal-space spectra $E_{\Lambda}({\bf k})$ in
Figs.~\ref{figch5:R47psilamk} and \ref{figch5:R104psilamk}(c) for
$n=4$ and $n=10$, respectively. One consequence of this is that the
order parameters $\langle{\hat{\Lambda}}_{\bf k}\rangle$, with ${\bf
  k}$ equal to the forcing wave vectors, do not vanish identically in
the disordered, turbulent phase; however, they do assume very small
values. Moreover, in the case of turbulence-induced melting the
crystal undergoes a transition from an ordered state to an undulating
crystal to a fully turbulent state.  Thus there is no noise and hence no
fluctuations in the crystalline state; i.e., it is equivalent to a
crystal at zero temperature. This has no analogue in the equilibrium
melting of a crystal.

In equilibrium, different ensembles are equivalent; we can, e.g., use
either the canonical or the grand-canonical ensemble to study the
statistical mechanics of a system and, in particular, the phase
transitions in it.  However, this equivalence cannot be taken for
granted when we consider nonequilibrium statistical steady states
(see, e.g., Ref.~\cite{ach00}).  We have seen an example of this in
Ref.~\cite{per09} where certain PDFs show slightly different
behaviours depending on whether we keep the Grashof number 
(i.e., the nondimensionalised amplitude of the force)
fixed or whether we keep the Reynolds number fixed.  Turbulence-induced melting
offers another example of the inequivalence of dynamical ensembles:
the precise sequence of transitions that we encounter in going from
the vortex crystal to the turbulent state depends on whether we do so
by changing the Grashof number as
in Ref.~\cite{bra97} or whether we do so by changing $\Omega$ as we
have done here. We have checked explicitly that we can reproduce the
sequence of transitions in Ref.~\cite{bra97} if we tune the Grashof
number rather than $\Omega$ to obtain the turbulent state.  

Investigations of similar transitions,
such as in the Kolmogorov flow~\cite{pla91,the92}, can 
benefit by using the combination of methods we have used
above. Detailed studies of the effects of confinement, air-drag induced
Ekman friction on turbulence-induced melting, initiated, e.g., in
Refs.~\cite{mol07,the92b}, can also make use of our
methods, but that lies beyond the scope of this paper. We hope,
therefore,
that our study will encourage experimental groups to analyse
turbulence-induced melting by using the set of techniques and ideas
that we have described above.

\section*{Acknowledgments}
We thank CSIR, UGC, and DST (India) for support, SERC (IISc) for computational 
resources, and D. Mitra, S.S. Ray, and K. Vijay Kumar for discussions. One of 
us (RP) is a member of the International Centre for Turbulence Research (ICTR).

\section*{References}

\end{document}